\documentclass[amssymb, amsmath, aip, apl, reprint]{revtex4-2}
\usepackage{graphicx}
\usepackage{amsmath}
\usepackage{amsfonts}
\usepackage{latexsym}
\usepackage{amsbsy}
\usepackage{epstopdf}
\usepackage{bm}%

\usepackage{color}

\usepackage{ulem} %for strikeout "\sout"

\newcommand{\dyadic}[1]{{#1}
\setbox0=\hbox{$\scriptstyle\leftrightarrow$}
   \setbox2=\hbox{$#1$}
   \dimen0=.5\wd0 \advance\dimen0 by-.5\wd2
   \advance\dimen0 by-\wd0
   \kern\dimen0
{^{\hbox{$\scriptstyle\leftrightarrow$}}}}

\begin{document}

\title{
{Independent Rydberg Atom Sensing using a Dual-Ladder Scheme}\\
}

\author{Samuel Berweger}
\author{Alexandra B. Artusio-Glimpse}
\author{Nikunjkumar Prajapati}
\author{Andrew P. Rotunno}
\affiliation{National Institute of Standards and Technology, Boulder,~CO~80305, USA}
\author{Noah~Schlossberger}
\affiliation{National Institute of Standards and Technology, Boulder,~CO~80305, USA}
\affiliation{Department of Physics, University of Colorado, Boulder, CO, 80309, USA}
\author{Dangka Shylla}
\affiliation{National Institute of Standards and Technology, Boulder,~CO~80305, USA}
\affiliation{Department of Physics, University of Colorado, Boulder, CO, 80309, USA}
\author{Kaitlin R. Moore}
\affiliation{SRI International, Princeton, NJ, 08540, USA}
\author{Matthew T. Simons}
\author{Christopher L. Holloway}
\affiliation{National Institute of Standards and Technology, Boulder,~CO~80305, USA}
\email{christopher.holloway@nist.gov}

\date{\today}

\begin{abstract}
Rydberg atom-based electric field sensing can provide all-optical readout of radio frequency fields in a dielectric environment.
However, because a single set of optical fields is typically used to prepare the Rydberg state and read out its response to RF fields, it is challenging to perform simultaneous and independent measurements of the RF field(s). 
Here we show that using two independent schemes to prepare and read out the same Rydberg state can be used to perform independent measurements in general, which we demonstrate specifically by resolving the the RF polarization.
We expect this work will be useful for fiber-coupled sensor heads where spatial multiplexing is challenging, as well as for complex multi-level sensing schemes.
\end{abstract}

\maketitle

Rydberg atom-based electric field sensing is an emerging technology that uses the interaction between optical and radio frequency (RF) electromagnetic fields with atomic vapors to enable all-optical detection of RF fields in a dielectric environment \cite{9748947,sedlacek2012,fan2015}.
In such a measurement, two or more optical beams are overlapped inside a vapor cell containing alkali metal (typically rubidium  or cesium) to prepare a Rydberg state and subsequently read out its RF response.
As such, the sensing volume is effectively the spatial extent of the overlapped beams within the vapor cell.
Because the optical fields are used to prepare and read out specific quantum mechanical states, simultaneous yet independent measurements beyond frequency multiplexing \cite{meyer2023} are not readily possible within the sensing volume. 
One possible approach for enabling entirely independent measurements within a single sensing volume is to generate Rydberg states by several excitation different paths (``ladders'') with no shared quantum coherence.

In this work we achieve two independent ladders using acousto-optic modulators (AOM) to generate beams that couple to the same Rydberg state using different hyperfine levels of the intermediate excited state. 
We spatially overlap these beams in a cross-polarized configuration within a vapor cell in order to perform simultaneous polarization-sensitive measurements.
We clearly show that the ladders respond independently to incident fields with polarizations aligned along or perpendicular to the polarization of the optical fields of each ladder.
In contrast, equivalent measurements performed with both optical fields using the same ladder no longer retain the capability to provide independent sensing.
We expect that this approach will prove useful for sensor head designs using optical fibers as well as loop schemes and other wave-mixing processes.

\begin{figure}%%%%%%%%%%%%%%%%%%%%%%%%
	\includegraphics[width=0.5\textwidth]{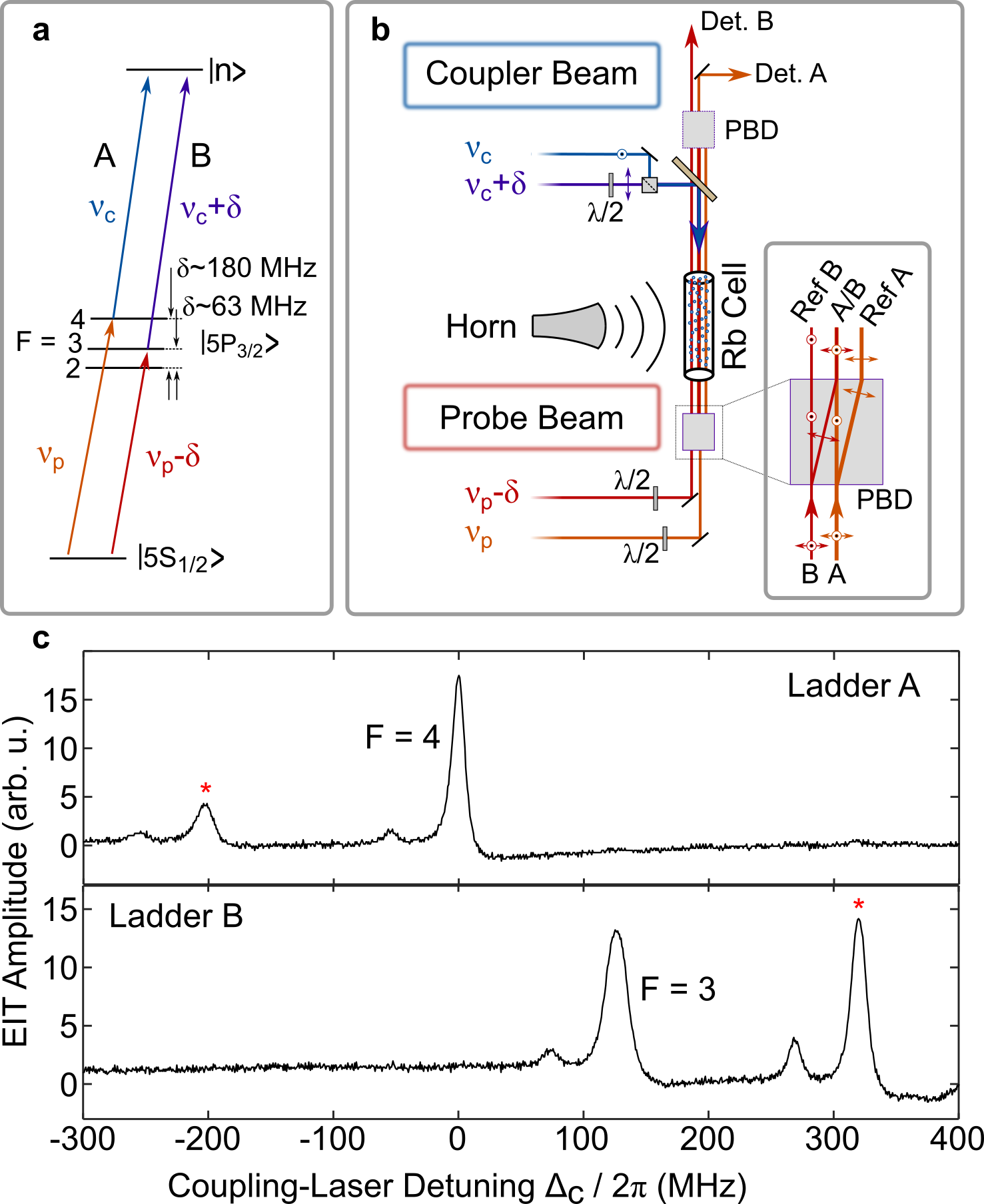}
	\caption{\textbf{Experimental Schematic.}} 
         (a) Illustration of the rubidium dual-ladder scheme used, with ladders A and B designed to use the F = 4 and F = 2 hyperfine states of the 5P$_{3/2}$ intermediate state, respectively.
         (b) Schematic of the experimental setup, which uses a combination of acousto-optic modulators (AOMs, not shown) and polarizing beam displacers (PBD) to generate the two ladders in a cross-polarized configuration at the same spatial position in the cell along with their respective reference beams.
         Shown in (c) are the simultaneously acquired EIT spectra for both ladders.
	\label{setup}
\end{figure}%%%%%%%%%%%%%%%%%%%%%%%%
Our approach uses rubidium-85 ($^{85}$Rb) atoms and is based on the two-photon scheme using electromagnetically induced transparency (EIT) that is commonly used for Rydberg atom electric field sensing \cite{shaffer2012,holl1, 9748947,fan2015}.
A linearly polarized ground state excitation (probe laser, 780~nm) is used to excite the $^{85}$Rb ground state (5S$_{1/2}$) to the first excited state (5P$_{3/2}$) along the D2 transition, while a second co-polarized counter-propagating field (coupling laser) then excites the atom from the 5P$_{3/2}$ state to the 60D$_{5/2}$ Rydberg state (480~nm).
As schematically illustrated in Fig.~\ref{setup}(a) the two independent ladders, denoted A and B, are designed to utilize two hyperfine levels of the 5P$_{3/2}$ state.
As illustrated in Fig.~\ref{setup}(b), AOMs are used to generate two beams from each laser that are shifted by $\delta$~$\approx$~180~--~200~MHz to produce two ladders that independently couple to different states. 
These beams are then combined using a polarizing beam displacer (PBD) to overlap the two ladders in a cross-polarized fashion at the same location in the cell while also producing two spatially separated reference beams.
After traversing the cell, the two beams are separated again using a PBD and independently detected using balanced detectors. 

We begin by examining the spectral characteristics of our dual-ladder scheme and their interactions. 
Shown in Fig.~\ref{setup}(c) are the EIT spectra simultaneously acquired for both ladders as the coupler laser frequency is swept across the 60D$_{5/2}$ Rydberg state.
The frequency axis is zeroed based on the position of the 60D$_{5/2}$ F~=~4 peak \cite{SteckRbData} of ladder~A, with the weaker 60D$_{3/2}$ seen 53~MHz lower in frequency.
The main peak seen for ladder~B is the F~=~3 peak at $\Delta_C$~=~125~MHz that arises from the doppler-shifts enabled by the large velocity distribution of our room-temperature vapor. 
Both the coupler and probe beams are detuned by $\Delta f_0$~$\approx$~83~MHz from the F~=~3 state, and the frequency shift is thus given by the ratio of the two wavelengths $\Delta f$~=~$\frac{\lambda_p}{\lambda_C}\cdot \Delta f_0$.
With the AOM set to 200~MHz, the F~=~2 peak is expected in ladder~B near the zero-frequency position, though it is weak and and is not readily discernible from any residual cross-talk from the primary peak of ladder~A.
% The F~=~2 peak for ladder~B is also shifted by a small amount due to the fact that our AOMs are operating at 200~MHz rather than the 180~MHz corresponding to the difference between the F~=~2 and F~=~4 levels.
%Both the coupler and probe beams are detuned by $\Delta f_0$~$\approx$~83~MHz from the F~=~3 state due to the 63~MHz offset between the F~=~3 and F~=~2 states as well as that both beams are $\approx$ 20~MHz detuned from the latter state due to the AOM frequency.

Due to the spatial overlap between the two ladders, additional peaks emerge due to the interaction between individual fields of the two ladders and their counterparts in the other ladder, i.e., the probe from ladder~B will see the coupler field from ladder~A and vice versa.
These peaks are indicated by an asterisk and can be identified by selectively blocking the individual coupling laser fields, or by modulating the AOM in the coupler beam path, which results in out-of-phase oscillation between the peaks corresponding to the ``pure'' ladders peaks and the peaks arising from cross-talk between the ladders. 
These peaks correspond to the F~=~4 probe transition of ladder~A seeing the coupler from ladder B, and the F~=~3 probe transition of ladder~B seeing the coupler from ladder~A.
These peaks are shifted by 200~MHz, as expected due to the frequency offset between the two coupler fields.
Because of the strength of the signal, we focus our efforts on the F~=~3 peak of ladder~B while noting that AOMs with an appropriate frequency shift (120~MHz) could be used to spectrally overlap the main peaks of the two ladders.

\begin{figure*}[!htb]%%%%%%%%%%%%%%%%%%%%%%%%
	\includegraphics[width=0.95\textwidth]{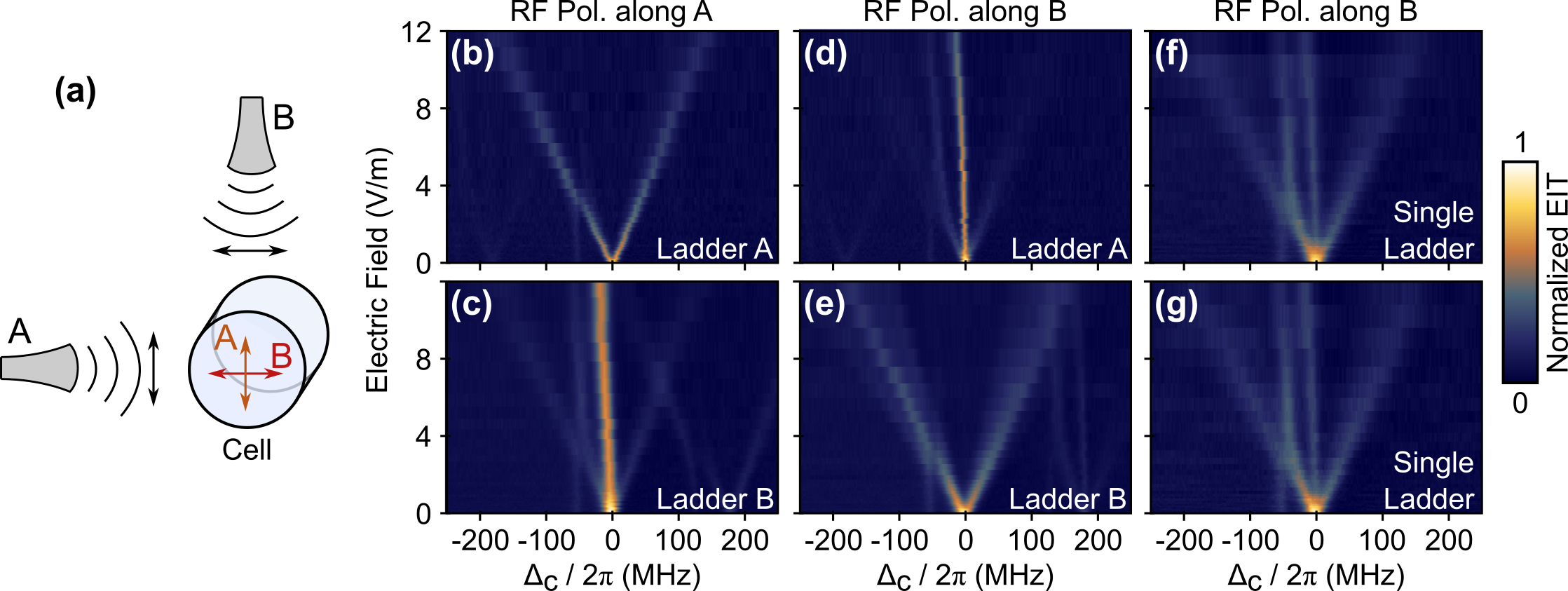}
	\caption{\textbf{Polarization-Resolved Detection.}
        (a) Schematic of the polarization discrimination with dual-ladders. Two horn antennas are positioned orthogonally such that each polarization is parallel with one of the optical field polarizations of either ladder~A or B.
        (b-c) False color plots showing the measured EIT spectra as the RF field strength was increased for ladders A and B as indicated with only Horn A on.
        (d-e) Same EIT spectra but with only Horn B on.
         (f-g) EIT spectra with only Horn B on, but only one ladder is used. Here (f) and (g) are the two readouts from the two different polarizations, but with no shift in the frequency (delta=0) between the two probes and couplers.
        } 
	\label{polarization}
\end{figure*}%%%%%%%%%%%%%%%%%%%%%%%%

In order to explore the utility of a dual ladder scheme for the independent sensing of two fields we examine two orthogonal RF polarizations. 
As schematically shown in Fig~\ref{polarization}(a), we position two horn antennas in an orthogonal configuration to generate RF fields at a frequency of 9.71~GHz to drive the transition from 60D$_{5/2}$ to 61P$_{3/2}$. 
The polarization of these RF fields is parallel with the optical fields of either ladder~A or ladder~B. 
Shown in Fig.~\ref{polarization} are the simultaneously acquired false-color plots of the evolution of the ladder~A (F~=~4)  and ladder~B (F~=~3) EIT peaks as a function of the applied RF field strength applied to ladder~A (b and c, respectively) and ladder~B (d and e, respectively).
These plots show distinctly different behavior for the two ladders.
For each applied RF field we see the characteristic Autler-Townes (AT) splitting expected for the corresponding ladder with all optical and RF fields co-polarized \cite{sedlacek2014}. 
However, the other ladder shows weak peaks with the same AT splitting but a significantly stronger residual main EIT peak that is unaffected by the RF field, which is also as expected when the optical and RF fields are orthogonally polarized \cite{sedlacek2014}.
This clearly shows that the two ladders respond to the RF fields independently and can be used for simultaneous sensing.

Lastly, we contrast our dual-ladder approach with a single-ladder approach.
In this case, we operate the system identically to the dual-ladder configuration, but the beams from both ladders are sourced from the same AOM output ($\delta$~=~0).
Shown in Fig.~\ref{polarization}(f,g) are the simultaneously acquired false-color plots from both single-ladder beams with the RF field applied along the ladder~B polarization. 
Here we clearly see that the responses are identical, showing that a single-ladder approach can no longer be decomposed into orthogonal polarization for independent measurements.

Our simple demonstration on polarization sensing reveals how a simultaneous and independent dual-ladder approach can be used to reduce a measurement that required active polarization rotation \cite{sedlacek2014} into a dual-channel measurement.
This may open up the path to improved transmission bandwith or independent sensing using polarization multiplexing. 
However, this approach can also be used for simultaneous sensing in general with, e.g, different RF transitions and is readily compatible with loop schemes \cite{berweger2023} implemented on each ladder.
More broadly, the use of independent hyperfine states can be used to independently link the ground state for six wave mixing and other complex loop schemes \cite{borowka2023,kumar2023}.
Although two ladders could be generated using separate lasers, the inherent simplicity of using AOM-generated sidebands limits the experimental complexity and cost.

The overall EIT peak height of the two ladders depends sensitively on the relative probe laser field strengths.
Although for weak probe fields the EIT peak height for either ladder increases as the respective probe power is increased, at higher powers increasing the probe power for one ladder will increase the corresponding signal while reducing that of the other ladder.
This is the result of the spatially overlapped beams that see the same atomic vapor and thus are competing for ground state atoms. 
This is the direct consequence of our desired effect, where the two ladders enable independent measurements by interrogating two separate atomic populations, which is not possible using a single ladder.
Since the combined sensitivity of the two ladders is limited by the ground state population, this could in principle be avoided by spatially separating the two fields.

The ability for independent sensing in overlapping beams holds several advantages.
First, spatial multiplexing presents significant challenges in fiber-coupled cells \cite{}, and the capability for simultaneous yet independent measurements using a single fiber significantly simplifies technical challenges.
Secondly, because of the influence of dielectric atom vapor cells on the spatial RF field distribution \cite{holloway2014,haoquan2015}, independent sampling of a single field should be performed at the same spatial location to enable direct comparison.
It should be emphasized that our choice of orthogonal polarizations for both ladders is one of convenience rather than necessity.
The orthogonal polarizations can be used to readily spatially separate and detect the two ladders, though separate detection could also be achieved based on the frequency separation of the two ladders using, e.g., optical heterodyne detection.

In this work we have implemented a dual-ladder scheme to independently generate Rydberg atoms within the sensing volume for RF field detection.
We reveal that this approach can readily be used for simultaneous and independent sensing, which we demonstrate by polarization-resolved detection. 
This approach is generally applicable and could find applications in fiber-based sensor heads as well more complex multi-state sensing schemes.

\textbf{Acknowledgements:} 
This research was developed with funding from the Defense Advanced Research Projects Agency (DARPA).
The views, opinions and/or findings expressed are those of the author and should not be interpreted as representing the official views or policies of the Department of Defense or the U.S. Government.
A contribution of the US government, not subject to copyright in the United States.

\bibliography{dual-ladder}

\end{document}